\documentclass[twoside,twocolumn,10pt]{article}
\usepackage[super,sort&compress,comma]{natbib} 
\usepackage[version=3]{mhchem}
\usepackage{graphicx} 
\usepackage{float}
\usepackage[english]{babel}
\usepackage{array}
\usepackage[T1]{fontenc}
\usepackage[usenames,dvipsnames]{xcolor}
\usepackage[normalem]{ulem}
\usepackage{amssymb}
\usepackage{amsmath}
\usepackage[left=1.5cm, right=1.5cm, top=1.785cm, bottom=2.0cm]{geometry}

\newcommand{\kpm}[0]{kcal~mol$^{-1}$}
\newcommand{\kpma}[0]{kcal~mol$^{-1}$~\AA$^{-1}$}
\newcommand{\icm}[0]{cm$^{-1}$}
\newcommand{\ala}[0]{\ce{Ala3+}}

\begin{document}

\title{\textbf{Machine Learning Molecular Dynamics\\ for the Simulation of Infrared Spectra}}
\date{}
\author{Michael Gastegger,\textit{$^{a}$} J\"org Behler,\textit{$^{b}$} and Philipp Marquetand\textit{$^{a}$}$^{\ast}$}

\twocolumn[
\begin{@twocolumnfalse}

\maketitle

\begin{abstract}
Machine learning has emerged as an invaluable tool in many research areas. In the present work, we harness this power to predict highly accurate molecular infrared spectra with unprecedented computational efficiency. To account for vibrational anharmonic and dynamical effects -- typically neglected by conventional quantum chemistry approaches -- we base our machine learning strategy on \emph{ab initio} molecular dynamics simulations. While these simulations are usually extremely time consuming even for small molecules, we overcome these limitations by leveraging the power of a variety of machine learning techniques, not only accelerating simulations by several orders of magnitude, but also greatly extending the size of systems that can be treated. To this end, we develop a molecular dipole moment model based on environment dependent neural network charges and combine it with the neural network potentials of Behler and Parrinello. 
Contrary to the prevalent big data philosophy, we are able to obtain very accurate machine learning models for the prediction of infrared spectra based on only a few hundreds of electronic structure reference points. This is made possible through the introduction of a fully automated sampling scheme and the use of molecular forces during neural network potential training. We demonstrate the power of our machine learning approach by applying it to model the infrared spectra of a methanol molecule, n-alkanes containing up to 200 atoms and the protonated alanine tripeptide, which at the same time represents the first application of machine learning techniques to simulate the dynamics of a peptide. In all these case studies we find excellent agreement between the infrared spectra predicted via machine learning models and the respective theoretical and experimental spectra.
\end{abstract}
\end{@twocolumnfalse} \vspace{3em}
]

\let\thefootnote\relax\footnotetext{\textit{$^{a}$~University of Vienna, Faculty of Chemistry, Department of Theoretical Chemistry, W\"ahringer Str. 17, 1090 Vienna, Austria.}}
\let\thefootnote\relax\footnotetext{\textit{$^{b}$~Universit\"at G\"ottingen, Insititut f\"ur Physikalische Chemie, Theoretische Chemie, Tammanstr. 6, 37077 G\"ottingen, Germany. }}
\let\thefootnote\relax\footnotetext{\textit{$^{*}$~E-mail: philipp.marquetand@univie.ac.at}}

\section{Introduction}

Machine learning (ML)  -- the science of autonomously learning complex relationships from data -- 
has experienced an immensely successful renaissance during the last decade.\cite{bishop_2006,Goodfellow-et-al-2016} Increasingly powerful ML algorithms form the basis of a wealth of fascinating applications, with image and speech recognition, search engines or even self-driving cars being only a few examples.
In a similar manner, ML based techniques have lead to several exciting developments in the field theoretical chemistry.\cite{Schuett2017,doi:10.1021/acscentsci.6b00219,1702.05532,Gomez-Bombarelli2016}

ML potentials are an excellent example for the benefits ML algorithms can offer when paired with theoretical chemistry methods.\cite{behler_neural_2011,handley_potential_2010,behler_representing_2014,Behler_2015_Per,doi:10.1021/acs.jpcc.6b10908,QUA:QUA24927,PhysRevLett.114.096405}
These potentials aim to accurately reproduce the potential energy surface (PES) of a chemical system (and its forces) based on a number of data points computed with quantum chemistry methods. Due to the powerful non-linear learning machines at their core, ML potentials are able to retain the accuracy of the underlying quantum chemical method, but are several orders of magnitude faster to evaluate. This combination of speed and accuracy is especially advantageous in situations where a large number of costly quantum chemical calculations would be required.

One such case is \emph{ab initio} molecular dynamics (AIMD), a simulation technique used to describe the evolution of chemical systems with time.\cite{Marx2012} In AIMD, the motion of the nuclei is described classically according to Newton's equations of motion\cite{newton1687philosophiae} and depends on the quantum mechanical force exerted by the electrons and nuclei.
AIMD is a highly versatile tool and has been used to model a variety of phenomena like photodynamical processes or the vibrational spectra of molecules.\cite{WCMS:WCMS64,B924048A,C3CP44302G,Mai2015IJQC,Marquetand_2016}

The latter application is of particular interest in the field of vibrational spectroscopy. With the development of more and more sophisticated experimental techniques, it is now possible to use methods like infrared (IR) and Raman spectroscopy to obtain highly accurate spectra of macromolecular systems (e.g. proteins).\cite{doi:10.1080/00268970903409812,IRHOBZ2007} As a consequence, vibrational spectra have become increasingly complex and theoretical chemistry simulations are now an indispensable aid in their interpretation.
Unfortunately, the standard approach to model vibrational spectra, static calculations based on the harmonic oscillator (HO) approximation, suffers from several inherent limitations.\cite{C3CP50739D,C3CP44302G}
Due to the HO approximation, anharmonic vibrational effects are neglected, which are of great importance in molecular systems with high degrees of flexibility and/or hydrogen bonding, such as biological systems. Moreover, HO based calculations are unable to account for conformational and dynamic effects, due to their restriction to one particular conformer. This also makes it hard to accurately model temperature effects, which have a large influence on conformational dynamics and are highly relevant for spectra recorded at room temperature.\cite{B924048A}
These deficiencies lead to disagreements between experimental and theoretical spectra, thus complicating a consistent analysis.

Different strategies, like the variational self-consistent field (VSCF) approach and its extensions\cite{C3CP50739D}, as well as quantum dynamics based methods\cite{WCMS:WCMS87,doi:10.1021/acs.jpca.5b10015}, have been developed to account for these effects, but they either neglect dynamical effects or are computationally intractable for systems containing more than a few tens of atoms.
Consequently, AIMD, which is able to describe anharmonicities, dynamic effects at manageable computational costs, is an invaluable tool for the practical simulation of vibrational spectra.\cite{B924048A,C3CP44302G}

Yet, standard AIMD is still comparatively expensive, placing severe restrictions on the maximum size of the systems under investigation (approximately 100 atoms) and on the quality of the quantum chemical method.
However, AIMD simulations can be accelerated significantly without sacrificing chemical accuracy by replacing the individual electronic structure calculations with much cheaper ML computations.
This opens the way for exciting new possibilities, making it possible to simulate larger systems and longer timescales in only a fraction of the original computer time.

The goal of the current work is to use ML accelerated AIMD calculations to simulate accurate IR spectra of different organic molecules.
This is achieved by harnessing the synergies between established techniques, improvements to existing schemes and new developments:
(I)   A special kind of ML potential, called high-dimensional neural network potential (HDNNP), is used to model the PES.\cite{behler_generalized_2007}  
(II)  Molecular forces are employed in the construction of these HDNNPs, using a novel method based on the element decoupled Kalman filter.\cite{gastegger_marquetand_2015}
(III) electronic structure reference data points are selected via an enhanced adaptive sampling scheme for molecular systems.
(IV)  A HDNNP based fragmentation method is used to accelerate reference computations for macromolecules.\cite{doi:10.1063/1.4950815} Finally, 
(V)   a new ML scheme to model dipole moments is introduced. 
A detailed description of all these individual components is given in the following section.

Three different molecular systems are studied using the strategies described above. First, a single methanol molecule serves as a test case to assess the overall accuracy of the HDNNP based simulations compared to spectra obtained with standard AIMD. 
Second, the ability of HDNNPs to efficiently deal with macromolecular systems is demonstrated by
(a) constructing a HDNNP of a simple alkane chain based only on small fragments of the macromolecule and (b) 
then using the resulting model to predict the IR spectra of alkanes of varying chain lengths.
In order to probe the suitability of HDNNPs for systems of biological relevance, a final study is dedicated to the protonated trialanine peptide. This also serves as an excellent test case for the ML based dipole moment model.

All HDNNPs are constructed using density functional theory (DFT) as electronic structure reference method.
Generalized gradient functionals are used in for methanol and the tripeptide.
In the case of alkanes, we demonstrate that in principle also highly accurate double-hybrid density functionals\cite{doi:10.1063/1.2148954} can be used.
The simulations carried out with these latter HDNNPs would be next to impossible using on-the-fly AIMD.
In all cases, comparisons to experimental IR spectra are shown.

\section{Theoretical Background}

In AIMD, vibrational spectra are computed via the Fourier transform of time autocorrelation functions.\cite{C3CP44302G} Different physical properties give rise to different types of spectra.
IR spectra depend on the molecular dipole moments:
\begin{equation}
I_\emph{IR} \propto \int^{+\infty}_{-\infty} \langle \dot{{\mu}}(\tau) \dot{\mu}(\tau+t) \rangle_\tau ~e^{-i\omega t} \mathrm{d}t,
\label{eq:dipauto}
\end{equation}
where $\dot{\mu}$ is the time derivative of the molecular dipole moment, $\omega$ is the vibrational frequency, $\tau$ is a time lag and $t$ is the time.

Upon closer examination of Equation~\ref{eq:dipauto}, several challenges to model AIMD quality IR spectra via ML become apparent:
Reliable ML potentials (and especially forces) are required to describe the time evolution of a chemical system.
Consequently, electronic structure reference points need to be selected from representative regions of the PES, while keeping the number of costly electronic structure calculations to a minimum.
This also calls for efficient strategies to handle the reference calculations of large molecules. 
And finally, a method to accurately model molecular dipole moments is required.

\subsection{High-Dimensional Neural Network Potentials.}

\begin{figure}[h]
\centering
  \includegraphics[height=4cm]{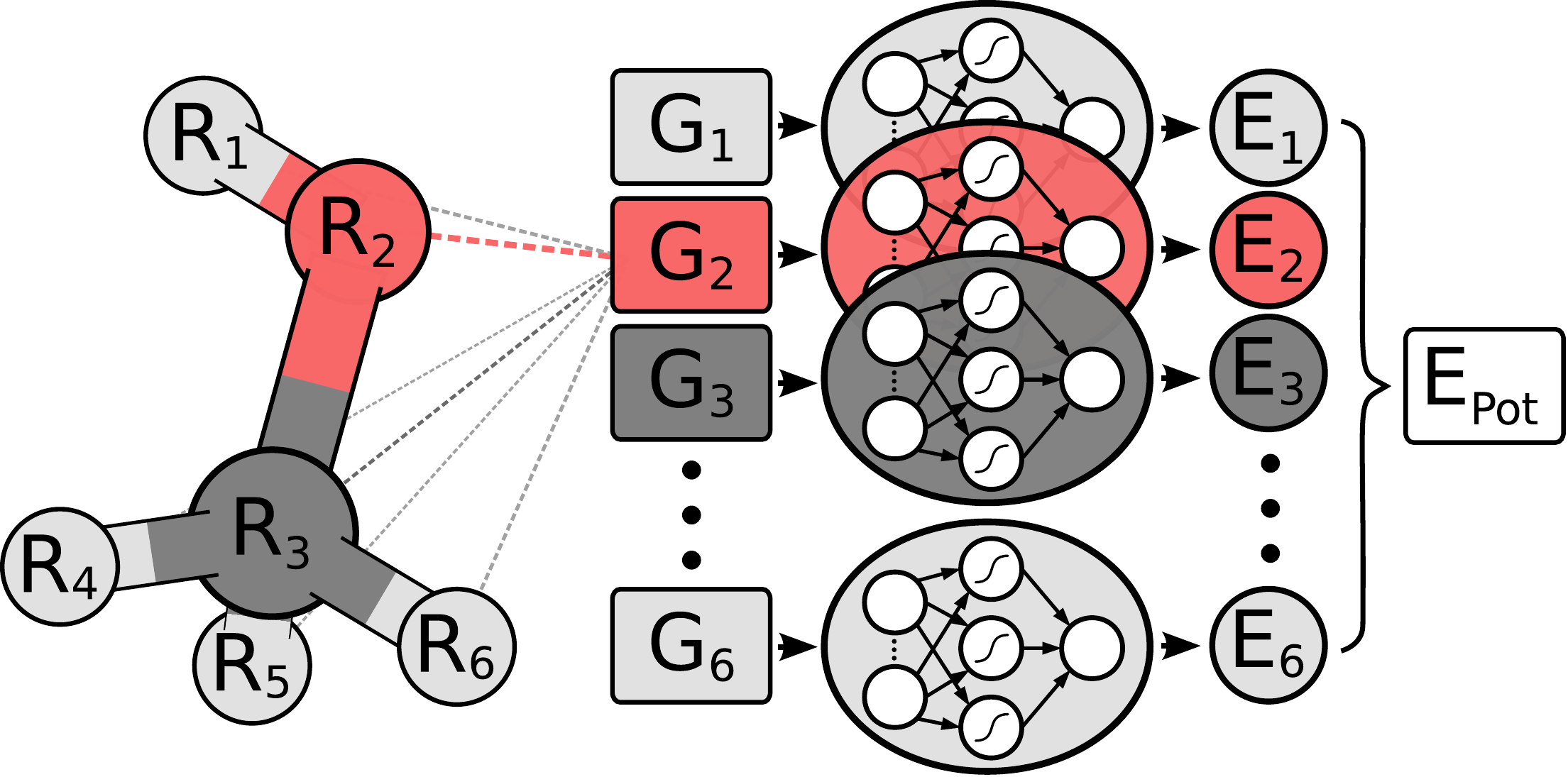}
  \caption{Schematic representation of a high-dimensional neural network potential (HDNNP). The Cartesian coordinates
  $\mathbf{R}$ are transformed into many-body symmetry functions $\{G_i\}$ describing an atom's chemical environment. Based on these functions, a NN then predicts the energy contribution $E_i$ associated with atom $i$. The potential energy $E_\mathrm{Pot}$ of the whole molecule is obtained by summing over all individual atomic energies.}
  \label{fgr:HDNNP}
\end{figure}

In a HDNNP (shown in Figure~\ref{fgr:HDNNP}), the total potential energy $E_\mathrm{pot}$ of a molecule is expressed as a sum of individual atomic energies.\cite{behler_generalized_2007,QUA:QUA24890}
The contribution $E_i$ of every atom depends on its local chemical environment and is modeled by a neural network (NN). These atomic NNs are typically constrained to be the same for a given element and thus, also termed elemental NNs.
Due to this unique structure, HDNNPs can easily adapt to molecules of different size and even be transferred between sufficiently similar molecular systems.

The chemical environment of an atom is represented by a set of many-body symmetry functions $\{G_i\}$, so-called atom-centered symmetry functions (ACSFs).\cite{behler_atom-centered_2011} 
ACSFs depend on the positions $\{R_i\}$ of all neighboring atoms around the central atom, up to a predefined cutoff radius.
By introducing a cutoff radius, an atom's environment is restricted to the chemically relevant regions. This brings two distinct advantages: the computational cost of HDNNPs now scales linearly with molecular size and chemical locality can be exploited in their construction and application\cite{behler_neural_2011}, which has been demonstrated recently e.g. for alkanes\cite{doi:10.1063/1.4950815}.
In addition, HDNNPs are well suited for molecular dynamics simulations, since an analytic expression for molecular forces is available due to their well-defined functional form.
For a detailed discussion of HDNNPs and ACSFs, see reference~\cite{QUA:QUA24890}. 

In order for HDNNPs to yield reliable models of the PES, a set of optimal parameters needs to be determined for the elemental NNs.
This is done in a process called training, where a cost function (typically the mean squared error) between reference data points (e.g. energies and forces) and the HDNNP predictions is minimized iteratively.
Different algorithms can be used to carry out the minimisation. The current work uses the element-decoupled Kalman filter\cite{gastegger_marquetand_2015}, a special adaptation of the global extended Kalman filter\cite{CEM:CEM1180080605} for HDNNPs.

Besides the energies, it is also possible to include molecular forces in the training process, by minimizing the cost function\cite{QUA:QUA24890}
\begin{equation}
\mathcal{C}_\mathrm{E,F} = \frac{1}{M} \sum^M_m \left( \tilde{E}_m - E_m \right)^2 + \frac{\eta}{M} \sum^M_m  \frac{1}{3N_m} \sum^{3N_m}_\alpha \left( \tilde{F}_{m\alpha} - F_{m\alpha} \right)^2.
\label{eq:pescost}
\end{equation}
The first term on the right hand side corresponds to the mean squared error between reference energies $E$ and HDNNP energies $\tilde{E}$. The second term describes the deviation between HDNNP ($\tilde{F}$) and quantum chemical forces ($F$). $M$ is the number of molecules in the reference data set, $N$ the number of atoms in a molecule, and $\alpha$ is an index running over the $3N$ Cartesian force components. $\eta$ is a constant used to tune the importance of the force error on the update step.
Including the forces in the training process leads to substantial improvements in the forces predicted by the HDNNP. 
Furthermore, instead of only one single energy, $3N$ points of additional information per molecule can now be utilized during training, thus greatly reducing the number of reference points required for a converged potential. An in-depth description of the element-decoupled Kalman filter and its extension to molecular forces can be found in reference~\cite{gastegger_marquetand_2015}.

\subsection{Adaptive Selection Scheme.}\label{sec:selection}

Ultimately, the quality of a ML potential does not only depend on the underlying ML algorithm and the employed training procedure, but also on how well the reference data set represents the chemical problem under investigation. Ideally, the reference data spans all relevant regions of the PES with as few data points as possible to avoid costly electronic structure computations.
To this end, different strategies -- e.g. based on Bayesian inference\cite{li_permutation_2013} or geometric fingerprints\cite{QUA:QUA24836} -- have been developed in the past.

\begin{figure}[h]
\centering
  \includegraphics[height=6.5cm]{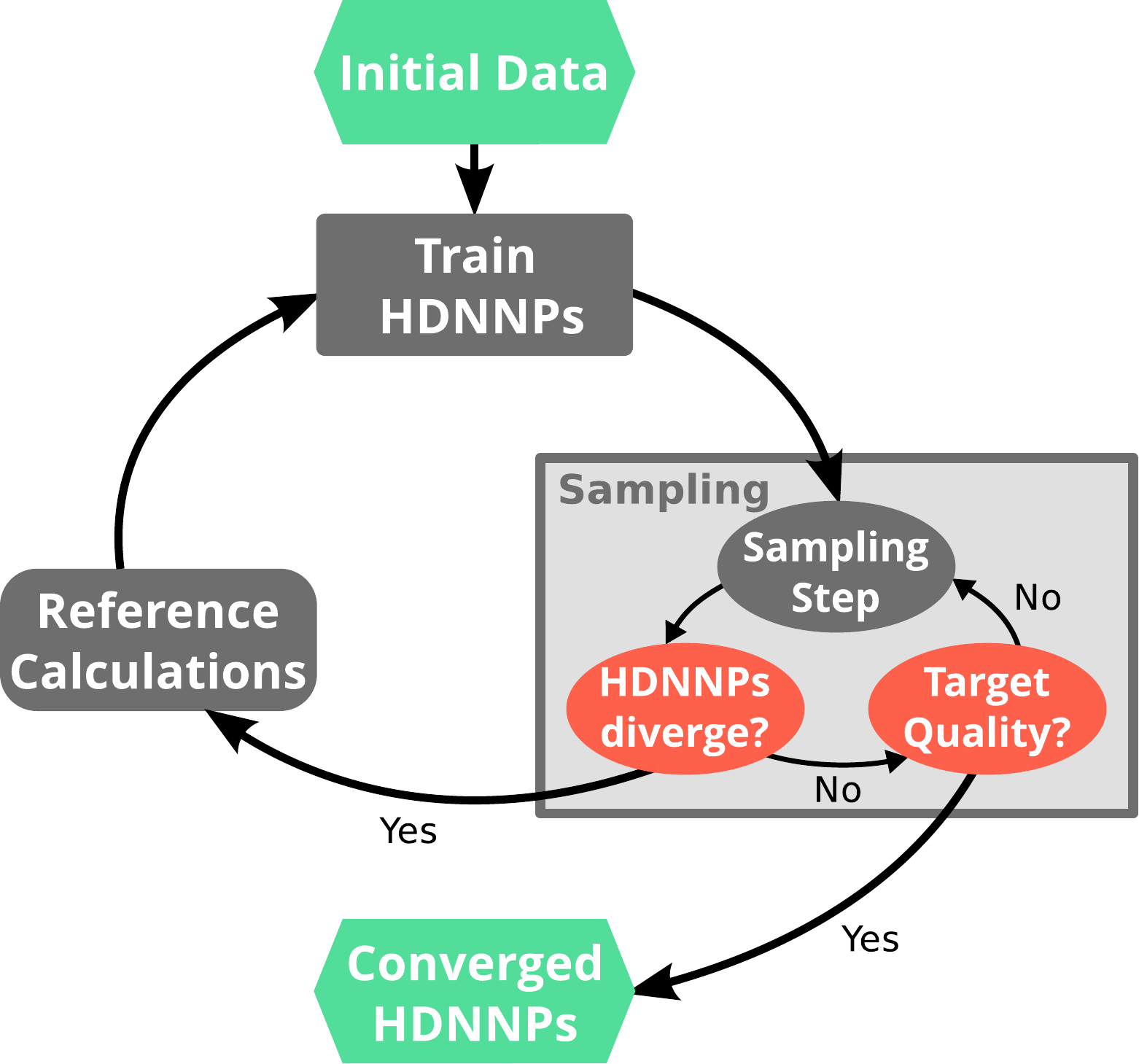}
  \caption{
A typical run of the adaptive selection scheme starts by using a small set of initial reference data points to train a preliminary ensemble of HDNNPs.
These HDNNPs are then used to sample new molecular conformations (e.g. via molecular dynamics simulations).
During sampling, the predictions of the individual potentials are monitored and if divergence is detected, the sampling run is stopped.
The conformation for which the HDNNPs disagree is computed with the electronic structure reference method and added to the set of reference points.
Subsequently, the HDNNP ensemble is retrained on the expanded data set and sampling is continued with the new potential.
This procedure is repeated in an iterative manner, until the divergence stops to exceed a predetermined threshold.
}
  \label{fgr:selection}
\end{figure}

A simple, but relatively effective procedure to select data points is based on the use of multiple HDNNPs and is described for example in reference~\cite{QUA:QUA24890}. 
After choosing an initial set of reference data points, a set of preliminary HDNNPs is trained, differing in the initial parameters and/or architectures of their elemental NNs (Figure~\ref{fgr:selection}).
These proto-potentials are then used sample different molecular conformations, using e.g. molecular dynamics simulations.
Afterwards, the predictions of the HDNNPs are compared to each other. Regions of the PES, where the different HDNNPs agree closely are assumed to be represented well, whereas conformations with diverging HDNNP predictions are modeled inaccurately.
The inaccurately described conformations are recomputed with the electronic structure reference method and added to the reference data set. The HDNNPs are then retrained using the expanded data set and the process is repeated in a self consistent manner until the HDNNPs reach the desired quality.

The current work introduces small adaptations to this procedure in order to make it more suitable for the use with biomolecules and expensive electronic structure reference methods.
Instead of performing independent sampling simulations with the individual HDNNPs, they are instead combined into an ensemble. In the ensemble, energy and forces are computed as the average of the $J$ different HDNNP predictions:
\begin{eqnarray}
\overline{E}          &= \frac{1}{J} \sum^J_{j=1} \tilde{E}_j,  \\
\overline{\mathbf{F}} &= \frac{1}{J} \sum^J_{j=1} \mathbf{\tilde{F}}_j. 
\label{eq:ensemble}
\end{eqnarray}
Simulations are then carried out using these averaged properties. The prediction uncertainty of the HDNNP ensembles is defined as
\begin{equation}
E_\sigma = \sqrt{ \frac{1}{J-1} \sum^J_j \left( \tilde{E}_j - \overline{E} \right)^2}.
\end{equation}
Ensembles of HDNNPs are less susceptible to erratic behavior in their individual parts. Moreover, the error of ensemble methods is typically proportional to $\frac{1}{\sqrt{J}}$, leading to a significant improvement in accuracy at negligible extra cost. The combination of both effects leads to more reliable simulations, especially in the early stages of PES exploration, hence diminishing the number of electronic structure starting points needed to seed the self-consistent refinement procedure. As a consequence, HDNNPs can now be grown on the fly from only a handful of data points in a highly automated manner: Starting from e.g. a few molecular dynamics steps, HDNNP ensemble simulations are run until $E_\sigma$ of a visited structure exceeds a predefined threshold. The corresponding conformation is recomputed with the reference method and added to the training set. The HDNNPs are retrained and simulations are continued from the problematic conformation. 

This procedure is effective, but highly sequential and calculations using expensive reference methods constitute a significant bottleneck.
Under the assumption, that the approximate shape of PES is sufficiently similar for different electronic structure methods, an ``upscaling'' step is introduced.
First, the iterative refinement is carried out using a low-level method until convergence of the HDNNPs. The conformations obtained in this manner are then recomputed using a high-level method. 
Since these high-level calculations can be done in parallel, the overall procedure is highly efficient.
Afterwards, new HDNNPs are trained, now at the quality of the better method.
The above assumption with regard to the similar shape of the PES at the different levels of theory is not necessarily valid, hence an upscaling step is typically followed by additional refinement steps at the higher level of theory.

A detailed discussion of the performance of the adaptive selection scheme and the convergence of the ML predictions with ensemble size can be found in the supporting information.

\subsection{Fragmentation with High-Dimensional Neural Network Potentials.}\label{sec:frag}

Since the computational cost of electronic structure calculations scale very unfavorably with system size and the accuracy of the underlying method, individual reference computations can still be problematic.
Hence, the required reference computations would quickly become intractable for highly accurate HDNNPs describing large molecular systems, despite the efficient sampling scheme.

It is possible to circumvent this problem by exploiting the special structure of HDNNPs.
As a consequence of expressing the HDNNP energy as a sum of atomic contributions and introducing a cutoff radius, HDNNPs operate the same manner as fragmentation methods using a divide and conquer approach:
Given only the energies of small molecular fragments, HDNNPs can reconstruct the energy of the total system.\cite{behler_neural_2011,doi:10.1063/1.4950815}
Thus, expensive electronic structure calculations never have to be performed for the whole molecule, but only for small parts of it.
The result is a linear scaling of the computational effort with system size.

\begin{figure}[h]
\centering
  \includegraphics[height=6.5cm]{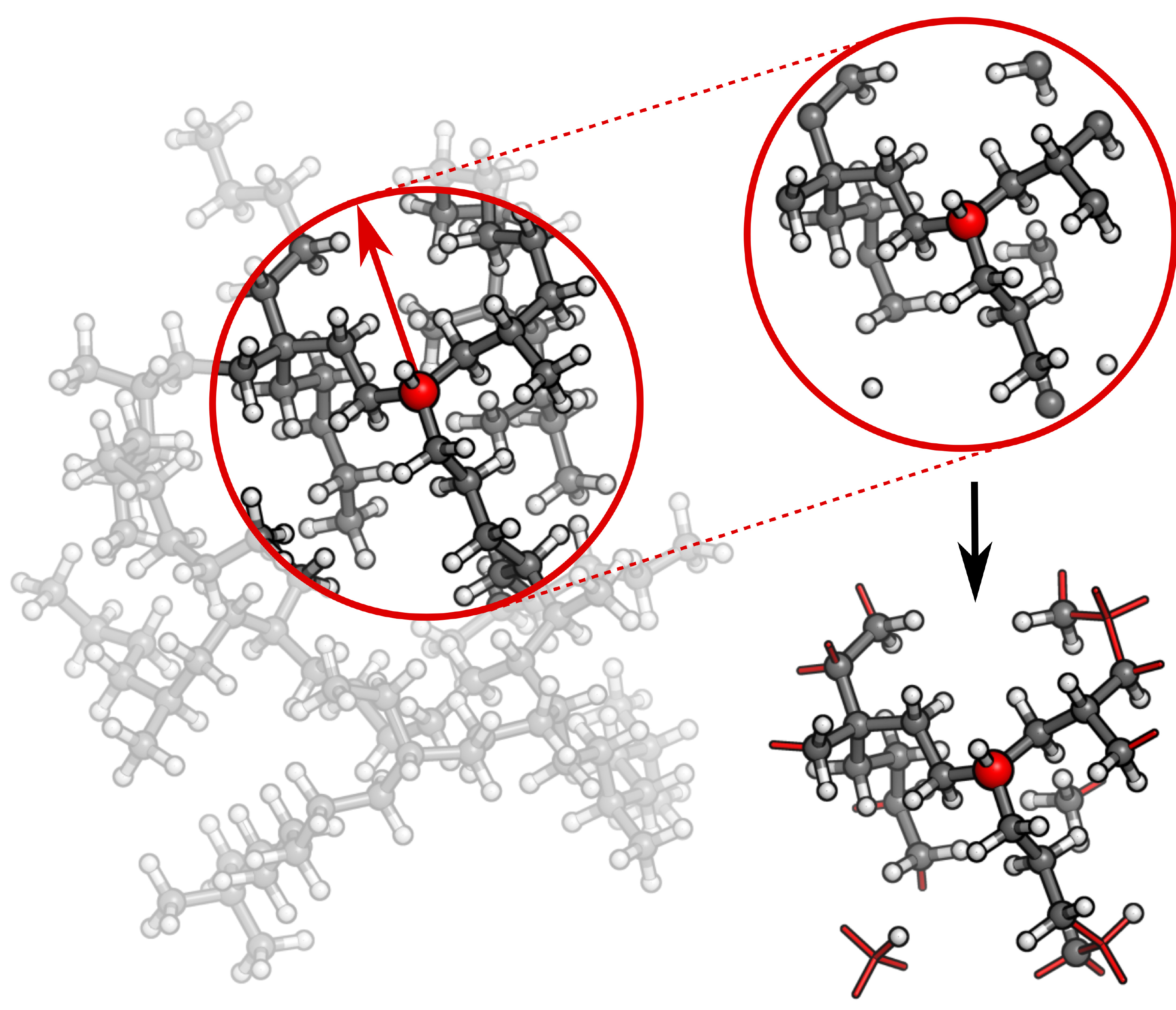}
  \caption{
In order to generate molecular fragments, first all atoms beyond a predetermined cutoff radius from the central atom are removed.
Afterwards, free valencies are saturated with hydrogen atoms, unless the valency itself is situated on a hydrogen or corresponds to a double bond in the unfragmented molecule.
In this case, the heavy atom connected to this atom in the original molecule is included in the fragment and the process is repeated iteratively.
This procedure is performed for the whole system, leading to one fragment per atom.}
  \label{fgr:fragmentation}
\end{figure}

In practice, a molecule is first divided into its individual fragments. Reference computations are then carried out for these fragments and the resulting data set is used to train a HDNNP. The ML potential is then applied to the geometry of the original molecule and the energy of the full system is recovered in this way.
Different strategies can be used to partition the full molecular system. In the current work, every molecule is split into $N$ atom-centered fragments (see Figure~\ref{fgr:fragmentation}). The size and shape of these fragments are determined by a cutoff radius around the central atom. Atoms beyond the cutoff radius are removed and free valencies are saturated with hydrogen atoms. 
If a free valency is situated on a hydrogen atom or two capping hydrogens overlap, the heavy atom corresponding to this position is instead included in the fragment and the process is repeated iteratively.
Typically, the same cutoff radius as in the ACSFs is used.

HDNNP fragmentation can easily be integrated into the adaptive sampling scheme. By using the deviations in atomic forces predicted by different HDNNPs as uncertainty measure, inaccurately modeled fragments can be identified. These fragments are then added to the reference data set.

\subsection{Neural Network Dipole Moments and Charge Analysis.}

A vital ingredient in the simulation of IR spectra with AIMD are molecular dipole moments (see Equation~\ref{eq:dipauto}).
While strategies to predict dipole moments using NNs exist\cite{doi:10.1021/ct800166r}, HDNNPs themselves have only been used to predict environment dependent charges in full analogy to the atomic energy contributions with the aim to model electrostatic long range interactions.\cite{artrith_high-dimensional_2011}

In this work, we extend this approach, by constructing molecular dipole moments as a sum of such environment dependent atomic partial charges:
\begin{equation}
\tilde{\mu} = \sum_i^N \tilde{q}_i \mathbf{r}_i,
\label{eq:dipmom}
\end{equation}
where $\tilde{q}_i$ is the charge of atom $i$ modeled by a NN and $\mathbf{r}_i$ is the distance vector of the atom from the molecule's center of mass.

While the elemental charge NNs could in principle be trained to reproduce charges computed with quantum chemical charge partitioning schemes (as was e.g. done in Reference~\cite{morawietz_neural_2012} to model electrostatic interactions), this approach has the following problems: 
First, the charge of a given atom obtained with such a partitioning scheme can in principle change along a trajectory in a non-continuous manner.
The resulting inconsistencies in the reference data can in turn lead to erratic predictions of the final model.
Second, unlike molecular energies and forces, atomic partial charges are no quantum mechanical observable. Hence, there is no physically unique way to determine them and a variety of different partitioning schemes exists.\cite{eltit} This complicates the choice of a suitable method to compute reference charges, since different schemes often exhibit vastly different behavior and sometimes fail to reproduce the molecular dipole moment accurately.\cite{JCC:JCC540141213}

Both problems can be avoided by training the elemental NNs to reproduce the molecular moments directly, while the environment dependent atomic charges $\tilde{q}_i$ are inferred in an indirect manner. In order to achieve this, a cost function of the form
\begin{equation}
\mathcal{C}_\mathrm{Q} = \frac{1}{M} \sum^M_m \left( \tilde{Q}_m - Q_m \right)^2 + \frac{1}{3M} \sum^M_m \sum^3_l \left( \tilde{\mu}_{lm} - \mu_{lm}\right)^2 + \ldots
\label{eq:QNN}
\end{equation}
is minimized. Here, $Q_m$ and $\mu_{lm}$ are the reference total charge and dipole moment components of molecule $m$. The index $l$ runs over the three Cartesian components of the dipole moment. $\tilde{Q}$ is the total charge of the composite NN model, computed as $\tilde{Q} = \sum_i^N \tilde{q}_i$, while $\tilde{\mu}$ is the NN dipole moment (Equation \ref{eq:dipmom}). While the cost function (from Equation~\ref{eq:QNN}) can be easily extended to include higher multipole moments, it was found that including only the total molecular charge and dipoles is sufficient for the purpose of modeling IR spectra. 
Since this scheme depends exclusively on molecular moments which are quantum mechanical observables, charge partitioning is no longer required. On the contrary, the trained NN model itself constitutes a new kind of partitioning scheme, where the atomic partial charges $q_i$ depend on the chemical environment and are determined on a purely statistical basis. These charges can also be used for additional purposes, e.g. to compute electrostatic interactions. Another possible application would be to augment classical force fields\cite{doi:10.1021/ct800166r}, where partial charges typically do not change with the chemical environment.\cite{mackerell_empirical_2004} As such, the NN charge scheme presented here constitutes an interesting 
alternative to static point charges or polarizable models.\cite{WCMS:WCMS1215}

\section{Computational Details}

Electronic structure reference calculations were carried out with \textsc{Orca}\cite{Neese2012} at the BP86\cite{B88,Dirac1929,Perdew1986,Vosko1980,Slater1951}/def2-SVP\cite{Weigend2005} (Methanol, \ala), BLYP\cite{B88,Dirac1929,Perdew1986,LYP}/def2-SVP (\ala) and B2PLYP\cite{doi:10.1063/1.2148954}/def2-TZVPP\cite{Weigend2005} (n-alkanes) levels of theory. All calculations were accelerated using the resolution of identity approximation.\cite{Eichkorn1995,Vahtras1993}

All HDNNPs were constructed and trained with the \textsc{RuNNer} program.\cite{RUNNER}
The NN dipole models were implemented in \texttt{python}\cite{Rossum} using the \texttt{numpy}\cite{Walt2011} and \texttt{theano}\cite{theano} packages.
Reference data points were obtained with the adaptive selection scheme, employing molecular dynamics trajectories at a temperature of 500~K with a 0.5~fs timestep to sample relevant conformations.
The final ML models are based on 245 (methanol), 534 (n-alkanes) and 718 (peptide) reference data points, with a maximum network size of 35-35-1 (two hidden layers with 35 nodes each and one node in the output layer) for the HDNNPs and 100-100-1 for the dipole moment model.

IR spectra were obtained with molecular dynamics simulations in the gas phase employing the same
timestep as the sampling procedure. After a short initial equilibration
period (3ps for methanol, 5ps otherwise), constant temperature molecular dynamics simulations were run for 30~ps in the case of methanol and 50~ps for the other molecules.
In addition to ML accelerated dynamics, AIMD simulations were carried out for methanol using the BP86 level
of theory described above.

Detailed information regarding the setup of the electronic structure calculations and molecular dynamics simulations, as well as the ML models can be found in the supporting information.

\section{Results and Discussion}

\subsection{Methanol.}

Due to its small size, the methanol molecule constitutes an excellent test system, not only for the direct comparison between IR spectra obtained via standard AIMD and ML simulations, but also to investigate the overall accuracy of the ML approximations.

\begin{figure}[h]
\centering
  \includegraphics[height=12cm]{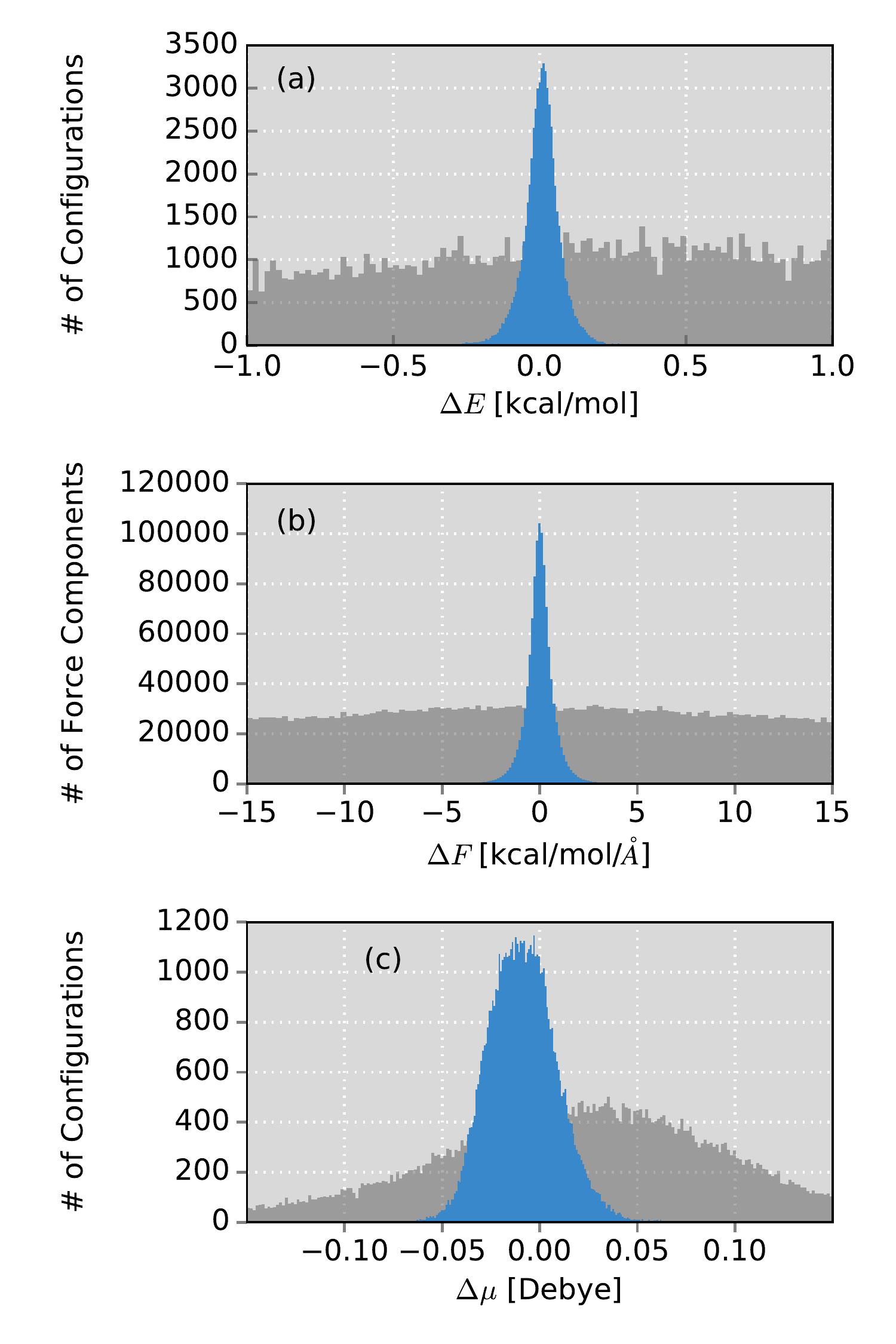}
  \caption{
Distribution of errors between the ML model based on the adaptive sampling scheme and the BP86 reference (blue).
The deviations were computed based on the electronic structure energies, forces and dipole moments (from top to bottom) of 60~000 configurations of methanol sampled with an AIMD simulation. The deviations obtained with a ML model trained on data points selected at random from a force field simulation are shown in grey (see supporting info).
}
  \label{fgr:errors}
\end{figure}

The final ML model for methanol consists of two HDNNPs and a NN dipole moment model trained on the BP86 data for 245 configurations.
To assess the errors associated with the individual components of the model, a standard AIMD simulation is run 30 ps, producing 60~000 configurations. 
For the sampled geometries, energies, forces and dipoles are predicted with the ML model.
These predictions are then compared to the respective electronic structure results.
The distribution of errors between ML predictions and the BP86 method are shown in blue in Figure~\ref{fgr:errors}.

Excellent agreement between electronic structure calculations and the ML model is found for all investigated properties.
In the case of energies (Figure~\ref{fgr:errors}a), the mean absolute error (MAE) of 0.048~\kpm\ (range of energies 13.620 \kpm) is well below the commonly accepted limit for chemical accuracy (1~\kpm) and is expected to be negligible compared to the intrinsic error of the electronic structure reference method in practical applications.
The components of the force vectors are reproduced equally well (Figure~\ref{fgr:errors}b), with a MAE of 0.533~\kpma\ (range 242.34~\kpma). These findings are comparable with other state of the art ML learning strategies developed specifically for the modeling of forces\cite{1611.04678} and demonstrate the excellent capabilities of HDNNPs to create potentials suitable for the dynamical simulation of molecules.
This conclusion is also supported by a comparison of the normal mode frequencies obtained for the optimized methanol structure at the ML- and BP86-level (see Table~\ref{tab:normalmodes}). 
\begin{table}[h]
\centering
\caption{
Comparison of the normal mode frequencies of methanol obtained with DFT and the ML model
}
\label{tab:normalmodes}
\begin{tabular}{crrr}
\hline
\# & DFT [cm$^{-1}$]& ML [cm$^{-1}$]& $\Delta$ [cm$^{-1}$]\\
\hline
1 & 331.70 & 346.94 & -15.24\\
2 & 1037.82 & 1030.00 & 7.82\\
3 & 1080.46 & 1092.09 & -11.63\\
4 & 1135.08 & 1138.21 & -3.13\\
5 & 1328.95 & 1320.84 &8.11\\
6 & 1420.02 & 1416.42 &3.60\\
7 & 1427.64 & 1422.59 &5.05\\
8 & 1449.79 & 1449.02 &0.77\\
9 & 2880.76 & 2892.94 &-12.18\\
10 & 2930.10 & 2961.48 &-31.38\\
11 & 3034.15 & 3054.08 &-19.93\\
12 & 3707.93 & 3707.73 &0.20\\
\hline
\end{tabular}
\end{table}
Although the HDNNP model was never explicitly trained to reproduce normal mode frequencies, its predictions agree well with the electronic structure frequencies, exhibiting a maximum deviation of only 31.38~\icm\ (0.090~\kpm).
The NN dipole model is also found to provide an accurate description of the molecular dipole moments (Figure~\ref{fgr:errors}c).
The total dipole moment shows an overall MAE of 0.016~D (over a range of 0.723~D) and the spatial orientation of the dipole vector is modeled equally reliable, with the MAEs of the individual Cartesian components ranging from 0.0173~D to 0.0200~D. The small shift of the dipole error distribution towards negative values is due to the fact that the atomic charges fluctuate around values other than zero. This effect is enhanced further, by the final summation to obtain the dipol moment model (see~\ref{eq:dipmom}).

\begin{figure}[h]
\centering
  \includegraphics[width=0.5\textwidth]{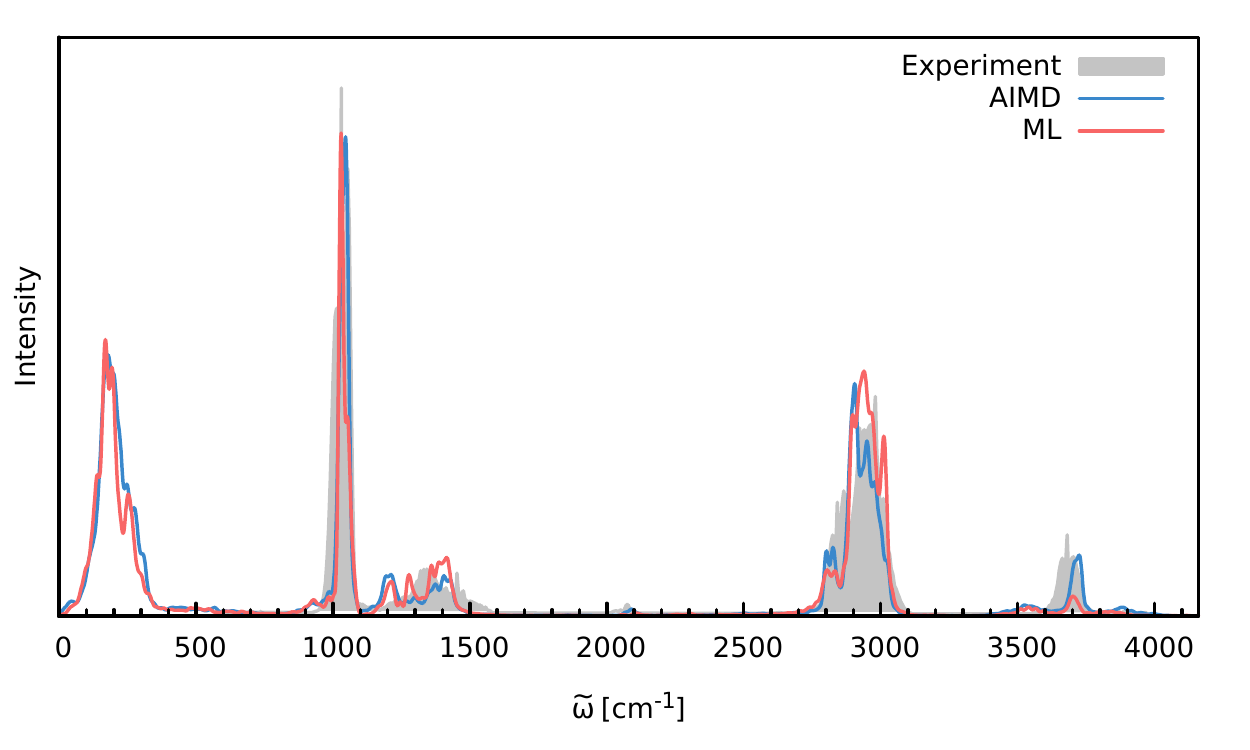}
  \caption{
IR spectra of the methanol molecule. The ML spectrum (red) is able to reproduce the AIMD spectrum (blue) obtained with BP86 with high accuracy.
In addition, both theoretical spectra agree well with the experimental one (grey).}
  \label{fgr:IRMeOH}
\end{figure}

In order to study the quality of the IR spectrum modeled with the composite ML model, it is compared directly to the spectrum obtained via the BP86 AIMD simulation.
Figure~\ref{fgr:IRMeOH} shows both IR spectra alongside an experimental spectrum of methanol recorded in the gas phase\cite{NIST}.
The overall shape of the ML spectrum, as well as the peak positions and intensities, show excellent agreement with the electronic structure reference.
The most distinctive difference between QM and ML spectra is the intensity of the stretching vibration of the O-H bond observed at 3700 \icm. This relatively minor deviation is most likely caused by small deviations of the dipole moment model.
Overall, the ML approach presented here is able to reproduce the AIMD IR spectrum of methanol with high accuracy. 
These results are remarkable insofar, as the final ML model is based on only 245 electronic structure calculations. This demonstrates the effectiveness of the combination of HDNNPs and the NN dipole model, as well as the power of the improved sampling scheme.

Finally, both simulations agree well with experiment, serving as an example for the utility of AIMD and ML-accelerated AIMD for the prediction of accurate vibrational spectra.

\subsection{n-Alkanes.}

\begin{figure}[h]
\centering
  \includegraphics[width=0.5\textwidth]{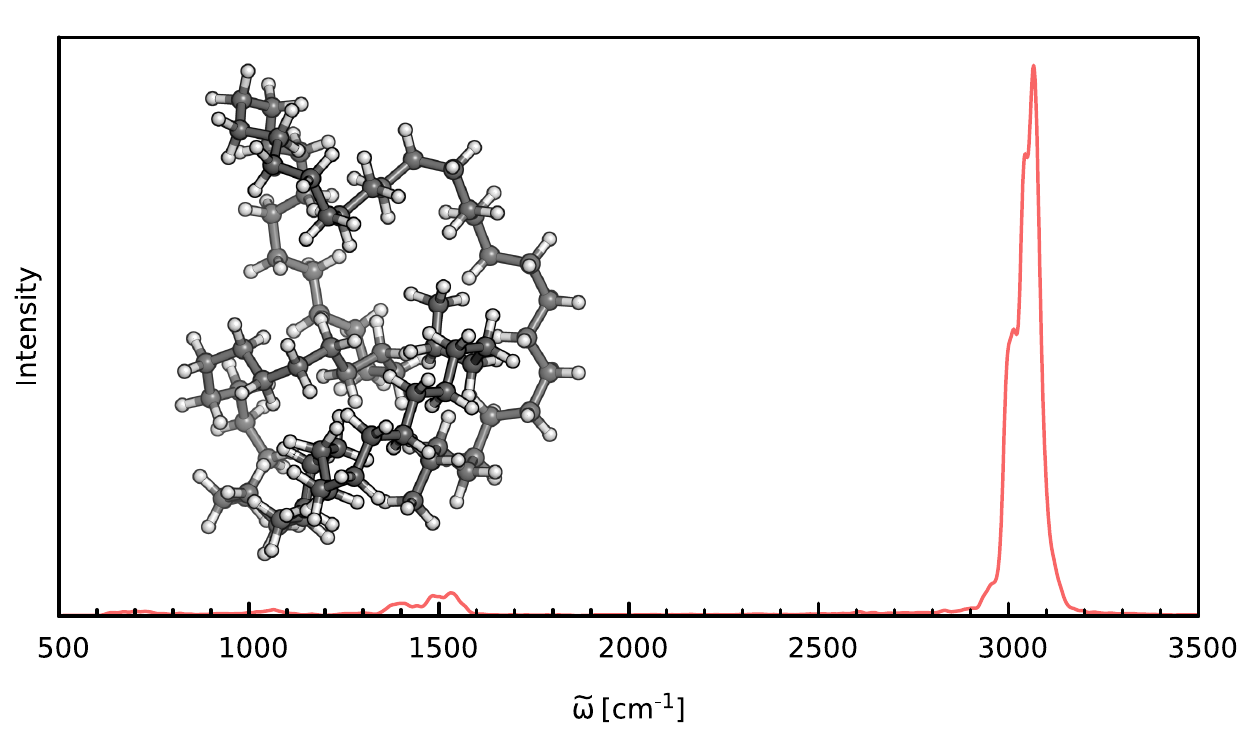}
  \caption{
IR spectrum of the \ce{C69H140} alkane as predicted by the ML model based on the B2PLYP method.}
  \label{fgr:IRAlk}
\end{figure}

When constructing ML potentials for large molecular systems containing hundreds or thousands of atoms, 
the necessary electronic structure reference calculations can quickly become intractable, especially for high-level electronic structure methods.
HDNNPs, as well as the dipole moment model presented in this work, can overcome this limitation via their implicit use of fragmentation (see Section ~\ref{sec:frag}).
In order to demonstrate the potential of this approach, it is used to predict the IR spectrum of an n-alkane with the chemical formula \ce{C69H140} (depicted in Figure~\ref{fgr:IRAlk})
via ML-accelerated AIMD simulations based on the B2PLYP double-hybrid density functional method.

The two HDNNPs and NN dipole moment model constituting the final ML model were trained on reference calculations for 534 fragments of the n-alkane.
These fragments use a cutoff radius of 4.0~\AA\ and contain 37 atoms on average and a maximum of 70 atoms. 
After initial adaptive sampling at the BP86/def2-SVP level, the final B2PLYP/TZVPP level ML-model is obtained via an upscaling step described in Section~\ref{sec:selection}.
Dispersion interactions, which are expected to play an important role in molecular systems of this size, are accounted for via a simple scheme: the HDNNPs are constructed from standard B2PLYP calculations and augmented with the empirical D3 dispersion correction using Becke--Johnson damping\cite{Grimme2010,BJDamp} in an \emph{a posteriori} fashion.

The IR spectrum of the \ce{C69H140} n-Alkane predicted via ML is shown in Figure~\ref{fgr:IRAlk}. It exhibits all spectroscopic features typical for simple hydrocarbons: The intense peak at 3000~\icm\ corresponds to symmetric and asymmetric C-H stretching vibrations. Deformations of the \ce{CH2}-groups give rise to the bands close to 1500~\icm, while the extremely weak signals in vicinity of 1000~\icm\ and 600~\icm\ are generated by C-C bond stretching and \ce{CH2} rocking vibrations.
Although the general shape and features of the IR spectrum are described well by the ML-model, some peak positions deviate from the expected experimental frequencies. This effect is especially pronounced for the C-H stretching vibrations, which are blue-shifted from the typical experimental value of 2900~\icm\ to 3040~\icm.

\begin{figure}[h]
\centering
  \includegraphics[width=0.5\textwidth]{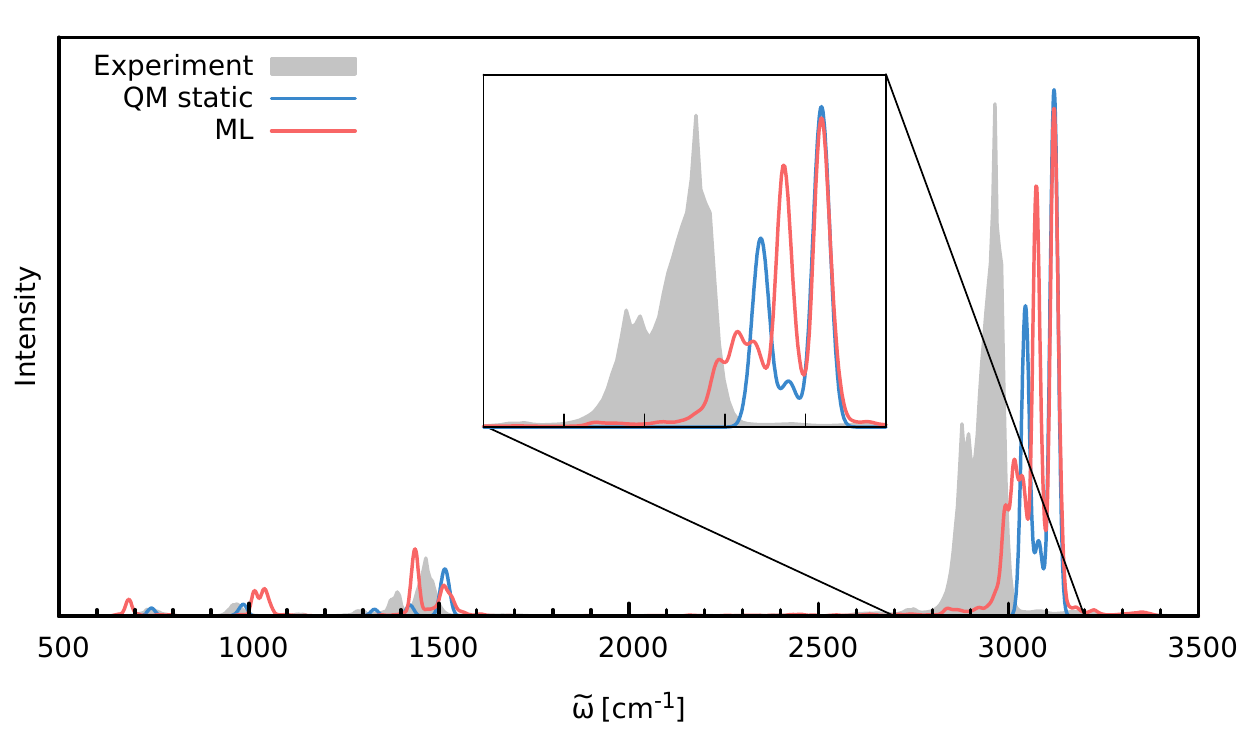}
  \caption{
IR spectrum of n-Butane obtained via the ML model (red), compared to the static quantum mechanical spectrum computed at the B2PLYP level (blue) and convoluted with Gaussians.
The peak positions in the ML and electronic structure spectra agree closely, suggesting that the observed deviations from experiment (grey) are due to the electronic structure method and not an artifact introduced by the ML approximation.
The overall structure of the peaks is reproduced much better by the ML accelerated AIMD simulation, especially in the region of the C-H stretching vibrations (see insert).}
  \label{fgr:IRBut}
\end{figure}

This blue shift is due to the electronic structure method (and not an artifact introduced by the ML approximations), as will be explained in the following.
Direct AIMD simulations and even static frequency calculations are prohibitively expensive for the \ce{C69H140} molecule. Instead, we exploit the transferability of the combined HDNNP and dipole model and use it to simulate the IR spectrum of the much smaller n-butane, for which static theoretical and experimental spectra can be obtained easily.
Figure~\ref{fgr:IRBut} shows the n-butane IR spectra obtained with ML-accelerated AIMD, static electronic structure calculations and experiment\cite{NIST}. 
The strong blue shift of the C-H stretching vibrations present in the ML spectrum can also be found in the static electronic structure spectrum.
Moreover, both spectra show good agreement with each other with respect to the overall positions of the spectral peaks.
These findings support the conclusion, that the observed frequency shifts are indeed a consequence of the underlying electronic structure method and not an artifact of the ML approximation.
Furthermore, the ML accelerated AIMD approach is found to accurately reproduce the structure of the experimental vibrational bands (especially the C-H stretching vibrations, see insert Figure~\ref{fgr:IRBut}).
This is not the case for the static spectrum and shows, that even for relatively small molecules an accurate description of dynamic effects is important in order to obtain high-quality IR spectra.
Both observations demonstrate the excellent accuracy of the HDNNP and NN dipole model, even for molecular systems not encountered during training.

Finally, to demonstrate the power the ML based approach in general and the fragmentation based approach in particular, a few exemplary timings are given for the \ce{C69H140} molecule (using a single core of an Intel Xeon E5-2650 v3 CPU):
Obtaining the relevant molecular fragments using the iterative sampling scheme takes approximately 7 days. The reference calculations of the fragments on the B2PLYP level of theory can be carried out in a highly parallel manner within 1.2 days (using a single CPU per configuration), including the time necessary to construct the final ML model. 
ML-accelerated AIMD simulations for the \ce{C69H140} molecule which involve the calculation of 110~000 energies and forces (5ps equilibration and 50ps simulation) take 3 hours. The NN dipole moments can be obtained within half an hour. Including the generation of the model, the total time to obtain the ML based IR spectrum amounts to a little over 8 days.
In contrast, the evaluation of a single energy and gradient at the B2PLYP level for the full n-alkane would require 30 days, extrapolating from the timings of the fragment reference calculations. Hence, performing the 110~000 calculations necessary for the AIMD simulation would require a total of 3.3 million days or 9~041 years.

\subsection{Protonated Alanine Tripeptide.}

Vibrational anharmonicities, as well as conformational and dynamic effects play a crucial role in the vibrational spectra of biomolecules.
In order to investigate the ability of ML accelerated AIMD to account for these effects, the composite ML model is used to simulate the IR spectrum of the protonated alanine tripeptide molecule  (\ala) in the gas phase.
Modeling the \ala\ molecule poses several challenges: 
An accurate description of the complicated PES depends crucially on the ability of the adaptive sampling scheme and the HDNNPs to reliably identify and interpolate relevant electronic structure data points.
Moreover, the changing charge distribution and dipole moment of the protonated species need to be captured by the NN dipole model.
Since the IR spectrum of \ala\ has been studied extensively, both experimentally and theoretically\cite{doi:10.1021/ct900057s,doi:10.1021/jp800069n}, the quality of the ML approach can be assessed directly.

The composite \ala\ ML model consists of two HDNNPs and a NN dipole model and was constructed from 658 reference geometries selected with the adaptive sampling scheme. 
The model exhibits overall RMSEs of 1.56 \kpm, 3.40 \kpma\ and 0.26 Debye for energies, forces and dipoles respectively.
This increase in the RMSEs and number of required data points compared to the previous systems is an indicator for the chemical complexity of the protein.
Long range dispersion interactions were accounted for in the same manner as in the case of the n-alkanes.

\begin{figure}[h]
\centering
  \includegraphics[width=0.45\textwidth]{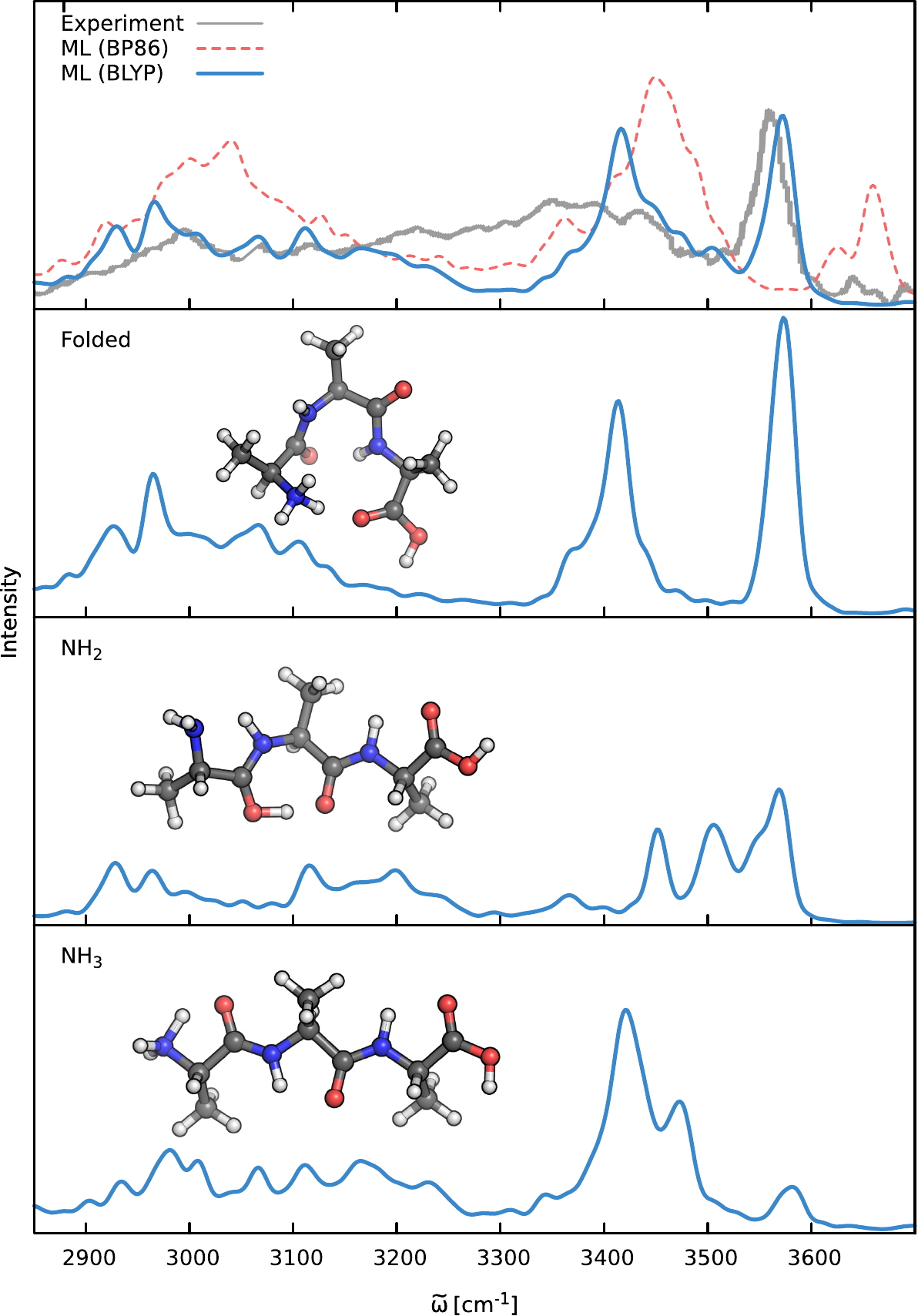}
  \caption{IR spectra of the protonated alanine tripeptide. The top panel shows the experimental spectrum (gray), as well as the ML spectra based on the BLYP (blue) and BP86 (red) reference methods. The lower panels depict the structures of the three main \ala\ conformers, along with their respective contributions to the averaged BYLP ML spectrum.}
  \label{fgr:IRTri}
\end{figure}

Previous theoretical studies by Vaden and coworkers\cite{doi:10.1021/jp800069n} have found, that the experimental IR spectrum of \ala\ is primarily composed of the contributions of three different conformers:
1) An elongated \ala\ chain with the proton situated at the N-terminal amine group,
2) a folded chain protonated at the same site and
3) a elongated form in which the proton is located a the carbonyl group of the N-terminus (see Figure~\ref{fgr:IRTri}),
which will be referred to as the \ce{NH3}, folded and \ce{NH2} families henceforth.
In order to account for these effects, ML accelerated AIMD simulations were carried out for all three conformers at 350~K, the estimated experimental temperature.
The final ML IR spectrum was then obtained by averaging. Figure~\ref{fgr:IRTri} shows the overall spectrum, as well as the contributions of the individual conformations alongside the experimental spectrum.\cite{doi:10.1021/jp800069n}
Due to the range of the recorded spectrum and the high congestion of spectral bands in the regions of the lower vibrational modes, we restrict our discussion only to the stretching modes involving hydrogens (ca. 2700~\icm\ to 3700~\icm).

As can be seen, the ML model correctly captures the features present in the experimental spectrum. The intense peak at 3570~\icm\ is due to the O-H stretching vibrations of the carboxylic acid group of the C-terminus. The position as well as the slight asymmetry of this band are almost perfectly reproduced in the ML spectrum. The region from 3300~\icm\ to 3500~\icm\ is populated by signals arising from the stretching modes of N-H bonds not participating in hydrogen bonds (e.g. \ce{NH2} terminus in the \ce{NH2} family). 
The free N-terminal N-H groups of the \ce{NH3} and folded family give rise to the intense feature at 3420~\icm. 
Vibrations associated with the N-H groups directly involved in hydrogen bonds are situated in the regions from 3100~\icm\ to 3300~\icm, where the ML spectrum captures several experimental subpeaks.
Finally, the region from 2800~\icm\ to 3100~\icm\ corresponds to the C-H stretching vibrations. Here, the most distinct features are the peak at 2930~\icm\ due to C-H vibrations of the C$_\alpha$ groups and the peak at 2970~\icm, which is caused by the vibrations of the methyl group hydrogens. 
The generally good agreement between the ML and experimental spectrum and the ability to reliably resolve individual bands is a testament for the efficacy of the composite ML scheme introduced in this work:
The dipole model is able to describe the charge distribution of \ala\ accurately, while the HDNNP ensemble provides a reliable approximate PES.

\begin{figure}[h]
\centering
  \includegraphics[width=0.45\textwidth]{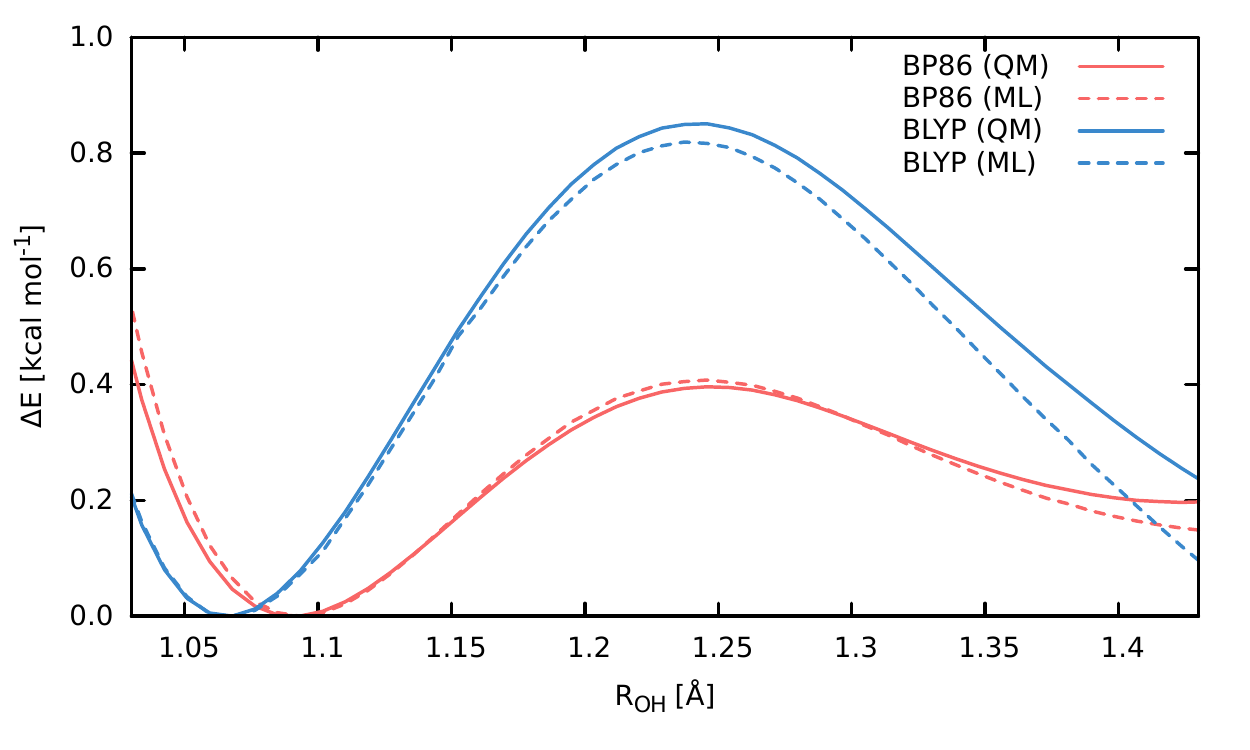}
  \caption{
Reaction barriers associated with the proton transfer from the N-terminal \ce{NH3} group in the \ce{NH3} conformer of \ala\ to the neighboring carbonyl.
The reaction coordinate is the distance between the transferred \ce{NH3} hydrogen and the carbonyl oxygen.
The barriers computed with the electronic structure reference methods are shown as solid lines colored red for the BYLP method and blue in case of BP86.
The dashed curves correspond to the predictions of the respective ML models, maintaining the above color scheme.}
\label{fgr:Htrans}
\end{figure}

A good perspective on the accuracy of the ML approach can also be gained by comparing the current ML model to one based on a different electronic structure reference method.
The top panel of Figure~\ref{fgr:IRTri} shows the averaged IR spectrum predicted by a ML model based on the BP86 density functional next to the previously discussed BLYP spectrum.
Although one would expect the closely related BLYP and BP86 methods to give similar results, significant differences can be found:
Besides a strong blue shift of the signal caused by the C-terminal COOH group by almost 80~\icm compared to the BLYP spectrum and experiment, large deviations are also found in the shape and positions of the bands corresponding to N-H stretching vibrations.
Here, we investigate the cause of the latter effect by closer examination of the \ce{NH3} conformer. Since the hydrogens of the N-terminal \ce{NH3} group can be involved in a proton transfer event to the neighboring carbonyl group, different spectra can arise depending on how often this transfer occurs. The transfer rate is directly correlated to the energy barrier associated with the transfer, suggesting that BLYP and BP86 differ significantly in the description of this event, which in turn leads to differences in the ML spectra.
Whether this phenomenon is caused by the ML approximations or due to the BP86 method itself, can easily be verified by computing the proton transfer barriers with both electronic structure methods and ML models.
As can be seen in Figure~\ref{fgr:Htrans}, the barrier height is indeed underestimated by the BP86 functional compared to BLYP, giving rise to the observed behavior.
At the same time, the ML models faithfully reproduce the barriers found with their respective electronic structure methods.
This is an excellent demonstration for the reliability of the ML approach, since the deviation between ML model and reference method is actually negligible compared to the differences between two closely related electronic structure methods.
The ease with which ML of different QM methods can be generated, also suggests a potential use of the ML approach presented here as an efficient tool for extensively comparing and thus benchmarking electronic structure methods.
Additional ML models can simply be constructed by recomputing in a parallel fashion the representative conformations selected by the sampling scheme with a different method and subsequent retraining of the new model (see Section\ref{sec:selection}).
Possible applications of this finding will be explored in the future.

The above observations also serve to highlight the ability of the ML model to automatically infer the chemistry underlying the \ala\ system.
Proton transfer events are essential in characterizing the experimental spectrum.\cite{doi:10.1021/ct900057s}
Driven by the automated sampling scheme, the composite ML approach gradually learns to describe these relevant chemical events, as is nicely demonstrated based on the reaction barrier previously obtained for the \ce{NH3} transfer (Figure~\ref{fgr:Htrans}):
Although the description of this event was never explicitly targeted in the training procedure, the barrier is nevertheless reproduced to an excellent degree of accuracy.
This feat is impressive insofar, as the ML model is based on an relatively small set of \emph{ab intio} computations.
These findings also serve to highlight an important advantage of HDNNPs over typical classical force fields, which is the ability to describe bond breaking and formation reactions.

Once again, the excellent computational efficiency of the composite ML model should be stressed:
While the computational chemistry method employed for \ala\ is already considered to be relatively cheap, the speedup gained is still significant.
A single step in the BP86 simulation takes approximately 1.5 minutes (on a single Intel Xeon E5-2650 v3 CPU).
The dynamics of every \ala\ conformer are simulated for 55~ps, requiring a total of 110~000 steps.
This amounts to a simulation time of 114 days for full AIMD. In contrast, using the ML model one can perform the same simulation in only one hour.

\section{Conclusions}

Here, we present the first application of machine learning (ML) techniques to the dynamical simulation of molecular infrared spectra.
We find that our ML approach is able to predict infrared spectra of various chemical systems in a highly reliable manner, correctly describing anharmonicities, as well as dynamic effects, such as proton transfer events.
The excellent accuracy -- which is only limited by the underlying computational chemistry method -- is paired with high computational efficiency, reducing the overall computation time by several orders of magnitude.
This makes it possible to treat molecular systems usually beyond the scope of standard electronic structure methods. As a proof of principle, we have simulated n-alkanes containing several hundreds of atoms, as well as the protonated alanine tripetide. However, even larger systems
can in principle be handled easily by our ML approach.
To realize the above simulations, we combined neural network potentials (NNPs) of the Behler--Parrinello type\cite{behler_generalized_2007} with a newly developed ML model for molecular dipole moments.
This neural network based model constitutes a new form of charge partitioning scheme based purely on statistical principles and offers access to environment dependent atomic charges.
For the efficient selection of electronic structure data points, a new adaptive sampling scheme is introduced. By employing this scheme, it is possible to incrementally grow ML potentials for specific applications in a highly automated manner based on only a small initial seed of reference data. 
When combined with the ability of NNPs to include molecular forces in their training procedure, the amount of electronic structure data points required to construct a ML potential is reduced tremendously (e.g. 700 conformations are sufficient for a converged potential of the tripeptide).
Furthermore, we demonstrate the ability of NNPs to model macromolecules based only on the information contained in small fragments, making it possible to treat even these systems with highly accurate electronic structure methods in a divide and conquer fashion.
The above findings are not only restricted to the simulation of infrared spectra via dynamics simulations, but apply to ML potentials in a broader sense. The ML approach presented here thus constitutes an alternative to the currently prevailing trend of fitting potentials to more and more reference data points. 
The latter strategy suffers from the disadvantage, that electronic structure reference calculations become prohibitively expensive for highly accurate methods and/or large molecular systems.
Here we show that these problems can be overcome through the efficient use of data,
bringing the dream of simulating the dynamics of e.g. enzymatic reactions with highly accurate methods one step closer.

\providecommand*{\mcitethebibliography}{\thebibliography}
\csname @ifundefined\endcsname{endmcitethebibliography}
{\let\endmcitethebibliography\endthebibliography}{}

\end{document}